# The seismogenic area in the lithosphere considered as an "Open Physical System". Its implications on some seismological aspects. Part - II. Maximum expected magnitude determination.

Thanassoulas, P.C., B.Sc in Physics, M.Sc – Ph.D in Applied Geophysics.

Retired from the Institute for Geology and Mineral Exploration (IGME)
Geophysical Department, Athens, Greece.
e-mail: thandin@otenet.gr – website: www.earthquakeprediction.gr

**Abstract.**

The (any) seismogenic area in the lithosphere is considered as an open physical system. Following its energy balance analysis earlier presented (Part – I, Thanassoulas 2008), the specific case when the seismogenic area is under normal (input energy equals released energy) seismogenic conditions is studied. In this case the cumulative seismic energy release is a linear time function. Starting from this linear function a method is postulated for the determination of the maximum expected magnitude of a future earthquake. The proposed method has been tested "a posteriori" on real EQs from the Greek territory, USA and data obtained from the seismological literature. The obtained results validate the methodology while an analysis is presented that justifies the obtained high degree of accuracy compared to the corresponding calculated EQ magnitudes with seismological methods.

## 1. Introduction.

There is a limited number of studies, in the Earthquake Prediction literature which refer to the magnitude determination of a future earthquake. The past seismic history (statistically treated) plus the dimensions of the studied, tectonic features of a seismogenic area provided some vague clues for the probable magnitude of the imminent strong EQ.

A physical model that will account for the determination of the prediction parameters of a strong EQ is, in a way, missing, from the seismological community.

The starting point of this topic is that different earthquake prediction parameters, most probably, require a different physical model which will be applied for its determination. Moreover, these physical models should not contradict but comply with each other. For the case of magnitude determination the "Lithospheric Seismic Energy Flow Model", which was introduced by Thanassoulas (2001, 2007, 2008), is used. This model is based on the balance of the absorbed and released energy, in the lithospheric seismogenic area, where an earthquake will occur in the future.

The rate of seismic energy release, as a function of an inverse power of the time, remaining to the main, seismic event, has been proposed and applied to back-analyses of several foreshock main sequences by Varnes (1987a, b).

In these papers, Varnes analyses the accelerating, precursory seismic activity in terms of seismic moment release. A simple, in time **(t),** empirical power law failure function was postulated (Bufe and Varnes, 1993) that related the parameters of the remaining time **($t_c$-t)** to the occurrence of the imminent, strong EQ and its corresponding magnitude **(M)** to the seismic moment release (the term accumulated Benioff strain is often used). Main (1995) studied the earthquakes from the point of view of a critical phenomenon.

In the same area of Statistical Physics (Main, 1996), Bowman et al. (1998) showed that the cumulative, seismic strain release increases, as a power law time to failure, before the final event. Papazachos et al. (2000) used the same methodology to study the Benioff strain rate release of the Aegean area. Moreover, an estimate for the timing of an imminent strong EQ was achieved with an accuracy of +/- 1.5 years, by using the same methodology for EQs of the Aegean area (Papazachos et al. 2001). The time to failure model was used, as a technique, in which a failure function is fitted to a time series of accumulated Benioff strain, by Di Giovambattista and Tyupkin (2004), in order to analyze the relation of the time-to-failure model to the hypothesis of fractal structure of seismicity. This method was applied on laboratory rock samples fracturing and real strong EQs of Kamchatka and Italy, as well.

In the early papers, that deal with the "Benioff strain - accelerated deformation – critical point" method, the studied seismogenic area was related to a circle of an optimum radius **R** and, lately, to an elliptical one (Papazachos et al. 2001).

The common feature, of all these papers, is the absence of a physical model that accounts for the well-known time to failure equation which is used very often. This equation is an empirical one and derives from the Statistical Physics, applied to other fields of applied science. Moreover, no tectonic information is taken into account, as far as it concerns the under study seismogenic area.

In this work, a new approach is used for the calculation of the magnitude of a future large EQ. It is based on the fact that the cumulative seismic energy release is a linear function of time when the under study seismogenic region is at a balanced energy flow state (Thanassoulas, 2008)

## 2. Theoretical model analysis.

It is well known that the stress, built up, in a focal area, is a very slow process which follows closely the motion of the lithospheric plates. It takes a long period of time (many years) to reach the point when an earthquake will occur because of the rock fracturing.

Under normal conditions, the stored energy is discharged through the normal (background) small magnitude seismicity of the area. When a strong earthquake is in preparation, this normal seismicity is decreased, for a certain time period, and therefore, the detected seismic "quiescence" was used as a precursory indicator.

The later is presented in the following figure **(1).** The cumulative number of seismic events, plotted, against time, deviates from the form of a linear function and resumes its normal one after the occurrence of the expected EQ.

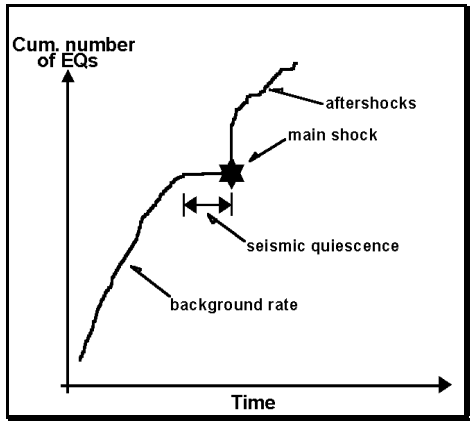

Fig. 1. Example of a seismic quiescence precursor.

Although this methodology gave successful results, with the way it treated the seismological data, it does not answer the question of the magnitude, neither the time of occurrence and it still remains in the domain of statistical methods. On the other hand, in terms of Physics, the same phenomenon can be studied from the energy transfer and balance point of view of the regional seismogenic area.

An open, physical system, that constantly absorbs energy, remains in a dynamic equilibrium when the in-flow energy is equal to out-flow energy, otherwise that physical system, should "explode", at a certain moment, releasing the excessive, stored energy, in a short time.

In the case of seismicity, the open physical system, in question, is the regional area of the lithosphere where the EQ will occur. Practically, the lithosphere is at a state of critical dynamic equilibrium, or it is teetering on the edge of instability, with no critical length scale (Bak, Tang and Wiesenfeld 1988; Bak 1996) as far as it concerns in-flow and out-flow of strain energy. The following figure **(2)** shows the postulated "energy - flow model" in such a condition.

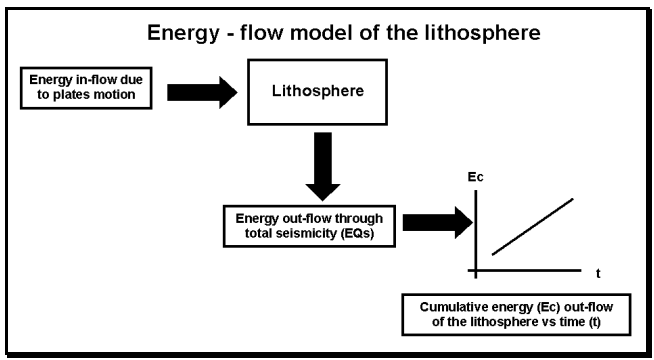

Fig. 2. Energy – flow model of the lithosphere and cumulative energy out-flow **(Ec)**, as a function of time.

Theoretically, following this working physical model, the cumulative energy release in a seismogenic area, during a normal, seismic period, should be a linear function (Thanassoulas, 2008) of time, due to energy conservation law of Physics. This is demonstrated in the lower right part of the figure **(2)**.

The cumulative energy **(Ec)** release function vs. time **(t)**, for a specific, seismogenic area, can be determined by converting the magnitude of each, recorded, EQ in the past, into energy, using any suitable formula that exists in the seismological literature. In the present work the following empirical formula (Tselentis, 1997) was used:

$$\text{LogE} = 11.8 + 1.5 M_s \tag{1}$$

Where:

**E** is the energy, calculated, in ergs.

**Ms** is the magnitude of each EQ, in Richter scale.

By applying this procedure to all EQs that occurred during the period of study, the cumulative energy Ec can be calculated and therefore, the graph **Ec = f(t)** can be constructed.

As an example for the expected results, over a seismogenic but of normal energy release period, the following figure **(3)** is presented. In a 20 years period of time that preceded the 1997/11/18, Ms 6.6R EQ in Greece, two distinct, normal energy release periods can be identified. The first one spans from 1 to 43, while the second spans from 67 to 205, in time scale (months). In these two periods the cumulative energy out-flow, due to normal seismicity, is that of the form of a linear function in time.

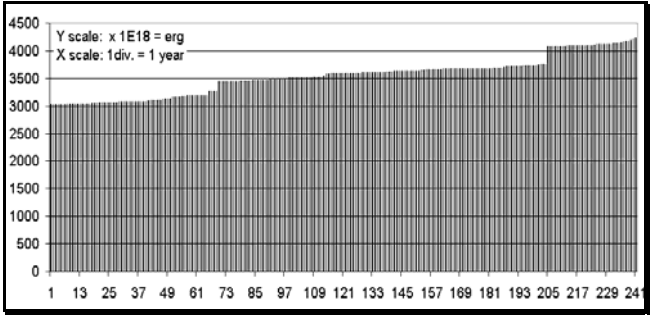

Fig. 3. Cumulative energy out-flow graph vs. time, for the seismogenic area of the EQ, on 18/11/1997 (Ms=6.6R) in Greece. Time period: 1977 – 1997.

Although this is the normal condition of the lithosphere, due to certain, mostly frictional, mechanical reasons, the seismogenic area, at a specific time, "locks". At this state the normal seismicity decreases and therefore the energy out-flow decreases. As an immediate result, the in-flow energy is stored in the seismogenic area. This stored excessive energy will be released in the future as an EQ of corresponding magnitude, at the time, when the seismogenic area "unlocks" during rock rupture. This is demonstrated in the subsequent figure (4).

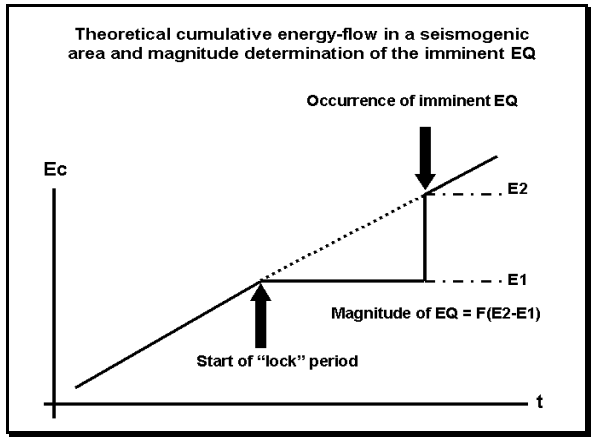

Fig. 4. Theoretical cumulative energy out-flow "lock" model.

In figure **(4),** the two black arrows indicate the start and the end of the "lock period". At the end of the "lock period", the imminent EQ takes place. Its magnitude is calculated by the two energy levels $E_{1,2}$ using formula **(1)**. After the expected EQ has occurred, the seismogenic region resumes its normal energy release rate, based on seismicity of small EQs.

A more general case is the one, in which the seismogenic area is always in critical strain conditions (mechanically locked) and earthquakes occur randomly in time, due to mainly frictional, mechanical reasons. This is demonstrated in the following figure **(4a)**.

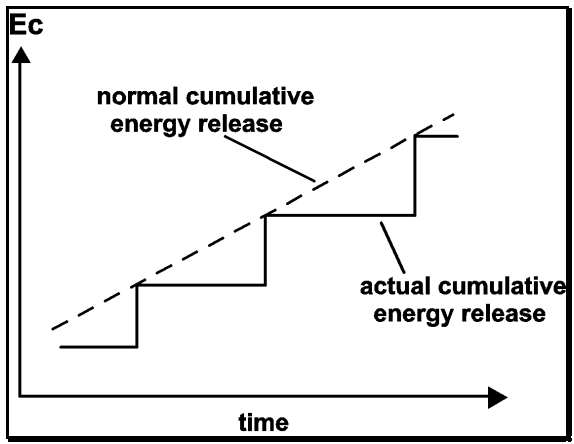

Fig. 4a. Step-wise cumulative, seismic energy release sketch presentation. The dashed line indicates the corresponding, normal, cumulative energy out-flow under stable conditions. Step-wise, solid black line represents the "always locked" mode of the lithosphere cumulative energy out-flow.

In such cases, the normal cumulative energy release time function is defined by the cumulative energy levels which were achieved at the time the earthquakes occurred. Therefore, this function is represented by the straight line (in least squares sense) that is defined by the cumulative energy steps (dashed line).

What is important, in this physical model, is the fact that, theoretically, during the lock period, the stored energy should equal the energy that could be released, due to the corresponding total EQs that didn't take place during the quiescence time period (energy conservation law of Physics).

The practical outcome of this is that, it is possible to calculate the maximum expected magnitude of the imminent EQ by knowing:

**a. The normal energy release rate of a seismogenic area.**

This is calculated from the past seismic history of the seismogenic area.

**b. The time of start of the lock period.**

This is determined from the graph **Ec = f(t),** where **(Ec)** is the cumulative energy of the seismogenic region, released, during the past period. This time is characterized by sudden decrease of the gradient **(dEc/dt)** of the **Ec = f(t).**

**c. The time of occurrence of the imminent EQ.**

The latter can be estimated, by other methods, very accurately (Thanassoulas et al., 2001)
It is obvious, from the graph **Ec(t),** that the longest the lock period is, the largest the magnitude of the EQ will be.
The application of this methodology has some prerequisites in order to be fulfilled.

**The first one** is a long, in time, seismic data history. This is required so that the normal out-flow energy rate, of the specific regional area, can be estimated, as correctly as possible. From this graph the start of the lock period can be determined, too.

**The second prerequisite** is the knowledge of the "end" of the "lock period". This is, generally, achieved by other methods (Thanassoulas et al. 2001). The monitoring of the change of various, other, physical parameters can signify the end of the "lock" period by the presence of their anomalous values. In this case, it is assumed that it was identified within an accuracy of +/- 1 month. Therefore, during the period of study, the **Ec** (cumulative energy) is determined at 1 month's sample time intervals.

The postulated methodology was applied, retrospectively, as an example, on a specific strong EQ in Greece. This EQ, having a magnitude of Ms = 6.6R, occurred in Greece on 1997/11/18, Lat. = $37.26^0$, Lon. = $20.49^0$.

In the following figure **(5)** the cumulative energy **(Ec)** is presented which was calculated for a spatial window $1.6^0 \times 1.6^0$ centered at the epicentral area and for the period from 1964 to 1997.

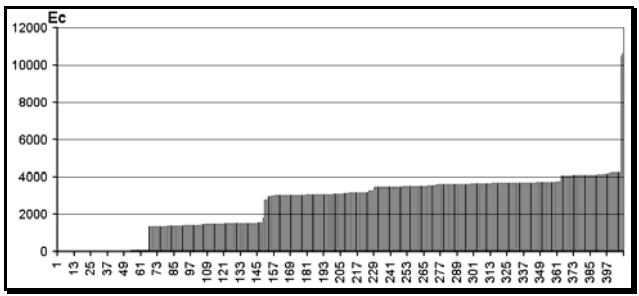

Fig. 5. Demonstration of a real case "locked" period. Sudden steps indicate the time of occurrence of EQs (abrupt energy release, Ecx1E18=ergs).

The **(Ec)** was calculated for a total period of 34 years, before its occurrence, with sampling interval of one month. Two major **(Ec)** increase steps, plus some smaller ones, characterize the graph. The step, located, on the right end of the graph, is the one that corresponds to the seismic event, its magnitude is to be, retrospectively, predicted. Major steps indicate large amounts of energy release (strong EQs), while smaller steps indicate the occurrence of seismic events of minor importance.

Consequently, taking into account short periods of time, for the seismic data history of the area under study, only small magnitude EQs can be estimated. On the other hand, taking into account long periods of time, it is possible to identify long, corresponding "lock" periods and therefore, large magnitudes can be calculated. In the present case, the rate of cumulative energy release was calculated by using the **(Ec)** levels, immediately, after the two strong events (67, 151 time axis of graph). The two cumulative energy levels, just after these two major, seismic events **(1969, 1976),** indicate the expected **Ec(t)** value under continuous normal seismic energy release **(Fig. 6)**.

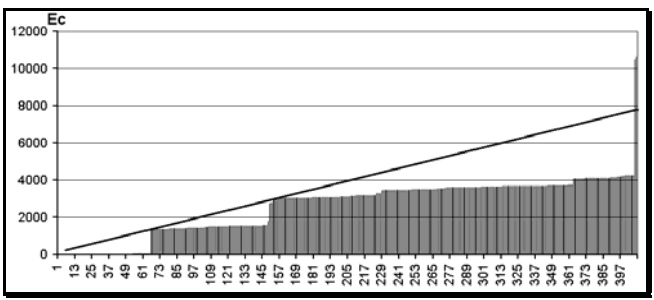

Fig. 6. Normal, out-flow energy (Ecx1E18=ergs) release rate (solid line) determination for the period 1964 - 1997.

The previous graph shows that prior to each strong EQ, a long "lock" period preceded it. The last "lock" period started on time 151, although seismic events, of minor importance, had occurred on 229 and 367 times. The later, released some energy, but not enough, to modify drastically the last identified "lock" period. In other words, the seismogenic area was only partially, discharged. It can be said that, the area under study was "locked" finally on time 151 (1976) and from that time on, it accumulated energy which was released through the 1997 strong EQ.

In the following figure **(7)** the determination of the magnitude is demonstrated.

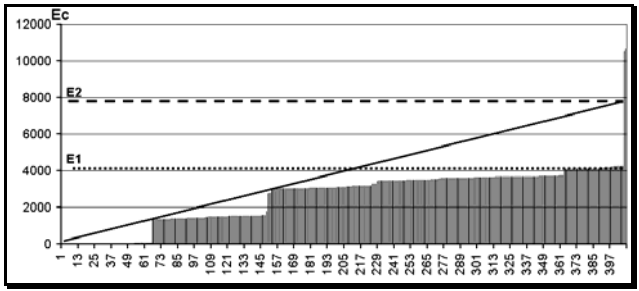

Fig. 7. Calculation, of the expected magnitude, assuming that the origin time of the imminent EQ is known. **(E1)** and **(E2)** lines indicate the initial and final cumulative energy levels. The solid line indicates the assumed cumulative energy out-flow-rate function, in time.

The **E2** level is determined from the intersection of a vertical line on the predicted time of the expected EQ and the normal out-flow energy release rate function.
Following this procedure, the calculated magnitude for the EQ on 1997/11/18 **(Ms = 6.6R)** was determined as **Ms = 6.5R**
The magnitude **Ms** which is calculated by this methodology deviates for only **.1 R** from what was determined **(Ms = 6.6R)** by the Geodynamic Institute of Athens.

**EQ: 18$^{th}$, Nov. 1997**   GEIN of Athens determination :   Ms = 6.6R
Energy-flow model application :   Ms = 6.5R

The same procedure was applied, retrospectively, for the determination of the studied by Mizoue et al. (1978) magnitude of EQs on Kii Peninsula, Central Japan. In this work the cumulative earthquake energy release of the under study area is presented in the following figure **(7a).**

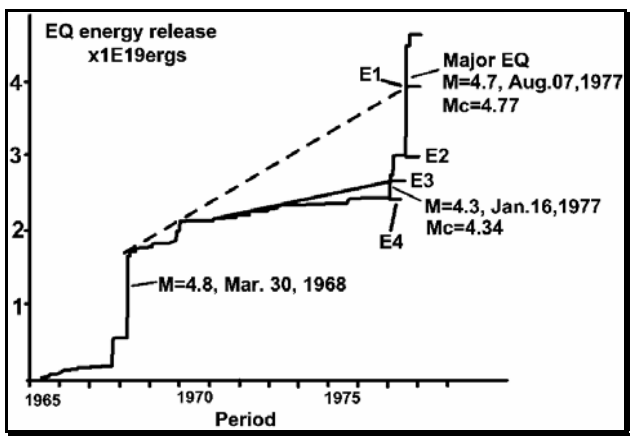

Fig. 7a. Cumulative energy earthquake energy release graph (after Mizoue et al. 1978). Energy levels E1, E2 resulted in a Mc=4.77, while E3, E4 resulted in a Mc=4.34 for the corresponding EQs. The dashed and solid lines represent the corresponding energy release rates for both cases.

The dashed line indicates the energy flow rate for the EQ on Aug. 07, 1977, M=4.7R, while the solid line indicates the energy flow rate for the EQ on Jan 16, 1977, M=4.3. The short, solid horizontal lines (E1, 2, 3, 4) indicate the minimum and maximum, cumulative energy levels which are taken into account for the calculation of the corresponding magnitude. In the case of the EQ on Jan 16, 1977, M=4.3R, the corresponding magnitude from fig. **(7a),** is calculated as M=4.34R, while for the EQ on Aug. 07, 1977, M=4.7, is M=4.77R.
In both cases, the error in the magnitude calculation is less than 0.1R, too.

**3. Statistical validation test of the methodology.**

A statistical test was applied to a larger data set (1972 - 2001 for EQs of Ms>=6.4R) so that the validity of the method could be verified. The data set was obtained by downloading the data file of Greek seismicity, for the period of time from 1964 to 2001, available on-line, by the Geodynamic Institute of Athens.

In this analysis the following parameters were adopted for the application of the postulated method:

**a. Time span:** The data file of the Geodynamic Institute of Athens extends (on line) back to 1964. Consequently, the seismic data were processed up to 1972 so that a minimum of 8 years past, seismic history would be available for the determination of the normal, cumulative energy out-flow rate for the subsequent strong seismic events to be studied.

**b. Spatial window:** It was selected as a $1.6^0 \times 1.6^0$ window, over the epicentral area to be studied, so that should be taken into account only the narrow, epicentral area, seismic history.

**c. Magnitude threshold**: This was selected as 6.4R so that only significantly large seismic events would be studied.

In the following **Table - 1** the 18 EQs, which were studied, are listed, ranking in magnitude scale. The first column indicates the year, month, day, hour, and the minute (yyyymmddhhmm format) when each event took place. The second column indicates the corresponding Latitude, the third column indicates the corresponding Longitude and the last one indicates its Magnitude in Ms **(Ms=M$_L$+.5)**.

**TABLE - 1**

| Date – time | Lat. | Lon. | Ms |
|---|---|---|---|
| 197904150619 | 42.00 | 19.00 | 7.3 |
| 198308061543 | 40.08 | 24.81 | 7.1 |
| 198201181927 | 39.90 | 24.50 | 6.9 |
| 198102242053 | 38.14 | 23.00 | 6.8 |
| 198112191410 | 39.20 | 25.30 | 6.8 |
| 198301171241 | 37.97 | 20.25 | 6.7 |
| 197205042140 | 35.30 | 23.60 | 6.6 |
| 199505130847 | 40.18 | 21.71 | 6.6 |
| 199607200000 | 36.11 | 27.52 | 6.6 |
| 199711181307 | 37.26 | 20.49 | 6.6 |
| 197806202003 | 40.80 | 23.30 | 6.5 |
| 198007090211 | 39.20 | 22.90 | 6.5 |
| 198112271739 | 38.90 | 24.90 | 6.5 |
| 197709112319 | 35.00 | 23.10 | 6.4 |
| 198102250235 | 38.20 | 23.00 | 6.4 |
| 198307051201 | 40.27 | 27.13 | 6.4 |
| 198406211043 | 35.36 | 23.31 | 6.4 |
| 199409011612 | 41.15 | 21.26 | 6.4 |

For each one of these EQs, the corresponding magnitude was determined, following the postulated methodology. The results are presented in the following **Table - 2**.

**TABLE - 2**

| Date – time | Observed Magnitude | Determined Magnitude | dM |
|---|---|---|---|
| 197904150619 | 7.3 | 7.25 | - 0.05 |
| 198308061543 | 7.1 | 7.06 | - 0.04 |
| 198201181927 | 6.9 | - | - |
| 198102242053 | 6.8 | 6.71 | +0.09 |
| 198112191410 | 6.8 | - | - |
| 198301171241 | 6.7 | 6.73 | +0.03 |
| 197205042140 | 6.6 | 6.40 | - 0.20 |
| 199505130847 | 6.6 | 6.15 | - 0.45 |
| 199607200000 | 6.6 | 6.51 | - 0.09 |
| 199711181307 | 6.6 | 6.59 | - 0.01 |
| 197806202003 | 6.5 | 6.35 | - 0.15 |
| 198007090211 | 6.5 | 6.51 | +0.01 |
| 198112271739 | 6.5 | 6.49 | - 0.01 |
| 197709112319 | 6.4 | 6.30 | - 0.10 |
| 198102250235 | 6.4 | 6.34 | - 0.06 |
| 198307051201 | 6.4 | 6.31 | - 0.09 |
| 198406211043 | 6.4 | 6.56 | +0.16 |
| 199409011612 | 6.4 | 6.29 | - 0.11 |

From a total of **18** strong EQs, for the **16** of them, it was possible to calculate the expected magnitude with a very high accuracy. This corresponds to an 89% success rate in magnitude calculation. For two of them the methodology failed. This is due to the fact that both EQs were large aftershocks of main events. Therefore, there was no past seismic history to be used for the application of the **LSEFM** methodology. Further statistical processing of the **dM** values results in:

**dM Mean = -.067R**

and

**dM S.Dev = 0.13R**

For the **11** out of the **16** EQs, for which the magnitude determination was possible, the dM value is less than **0.1R**, corresponding to a **73.33%**. One only extreme value of –0.45 was calculated and the rest **4** of them were calculated with a **dM** value, ranging between **0.1R** and **0.2R**.

### 4. Explanation of the achieved accuracy.

The accuracy, in the calculation of the magnitude for a pending, future earthquake, is surprisingly large. Whenever this methodology was discussed with seismologists, it was difficult for them to accept such a small deviation from the magnitudes, calculated by traditional, statistical, seismological methods. It must be pointed out, too, that the predicted, seismological magnitudes are accepted that deviate at least +/- 0.5 R from the real ones.

Let us assume that a seismological group calculates magnitudes of seismic events. The calculated magnitudes are affected by the error tendency of the group, due to various causes. The introduced errors on the magnitude calculations are represented by a "white noise", superimposed, on the real magnitude values, since it is a random process.

This process is presented in the following figure **(8)**.

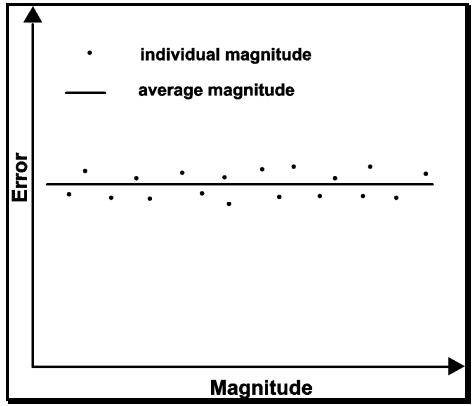

Fig. 8. The dependence of introduced errors (black dots), by a seismological group, in the calculation of the real magnitude value (solid horizontal line) of a seismic event is presented in sketch-drawing.

The real magnitude value, of each calculated seismic event, will correspond with the one indicated by the determined, straight line, in a least square sense, through all the individual calculations made by the same, seismological group.

If these "noisy" seismic magnitudes are used to calculate the cumulative, seismic energy release, then the following graph of figure **(9)** results. The black part of the graph corresponds to the past seismic events. In the case a future seismic event takes place, the stored, cumulative, seismic energy that will be released and which corresponds to that time of the seismic event, will be located in the extrapolated (red) part of the past (black) graph. This is presented in the following figure **(9)**.

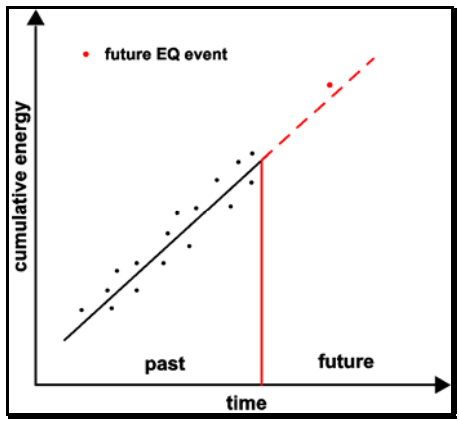

Fig. 9. Sketch presentation of the cumulative energy that corresponds to the past (black) and future (red), seismic events.

Consequently, in case of a future earthquake (red dot) of which the time of occurrence is already known by other means, the stored, cumulative energy that corresponds to its time of occurrence, can be found by extrapolating the cumulative energy black graph of figure **(9),** in time. Therefore, working backwards for the calculation of its magnitude, inevitably, its magnitude will deviate at most, as indicates the "white noise", which was introduced by the seismological group which made the calculations. Even better, the deviation will be less, since it will be located on the real value magnitude graph which is calculated in the least squares sense. The later is presented in the following figure **(10)**.

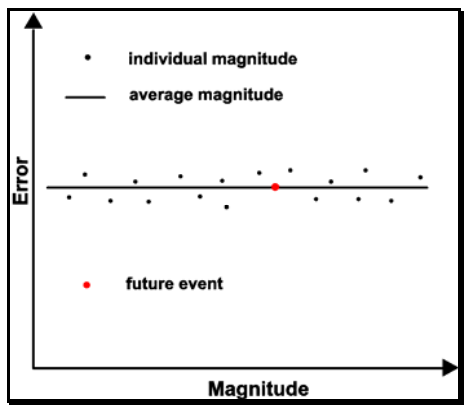

Fig. 10. Correlation of the magnitude of a future, seismic event (red dot) in relation to magnitudes (black dots), calculated, by seismological methods and the errors involved in this seismological procedure.

In conclusion, the achievement of better accuracies (in the majority less than +/- 0.1 R) in predicted magnitudes, compared with the calculated ones (+/- 0.5 R at best) by statistical seismological methods, should not be considered as a surprising fact.

Actually, what can be said in simple words, about this technique is that, **past seismicity constructs the cumulative energy graph in a least-square sense and this very same graph is exactly used to calculate, backwards, the exact magnitude of the future, seismic event.**

## 5. Examples of the application of the methodology on real EQs.

The "Lithospheric Seismic Energy Flow Model" methodology was tested "a posteriori" against known past strong earthquakes, whose seismic parameters of location, time of occurrence and seismological magnitude are already known. The following earthquakes were selected as typical examples:

a. Skyros, Greece, 26/7/2001, Ms = 6.1 R.

b. Northridge EQ, California, USA, 17/1/1994, M = 6.7 R.

c. Parkfield EQ, California, USA, 28/9/2004, Mw = 6.0 R.

The last two examples were chosen in order to make it clear that, the methodology is undependable on place and seismology research group, provided that a regular, earthquake catalog is available, concerning the past seismicity of the under study area.

### 5.a. Skyros, Greece, 26$^{th}$ July 2001, Ms = 6.1 R.

The postulated methodology was applied on Skyros EQ. A few days before the occurrence of Skyros EQ, electrical precursory signals were recorded by the (**VOL**) monitoring site.
The location of Skyros azimuthal direction resulted as the source of their origin after detailed processing of these signals (Thanassoulas et al. 2001). Moreover, suddenly the seismic activity increased on Skyros regional area (EQ on 21$^{st}$ July, Ms = 5.1R took place), indicating thus, in an indirect way, the regional area that was seismically activated and consequently generated the emitted and recorded, electrical signals.
Starting with this hint, the postulated methodology was applied to a window of **1.6$^0$x1.6$^0$**, centered, on the epicentral area of the 5.1R EQ on **21$^{st}$ July**, in an attempt to predict the magnitude of the expected EQ. The timing of Skyros EQ was calculated by other methodology (Thanassoulas et al. 2001, 2001b). Therefore, it was a simple task to detect the presence of any seismic "lock" state of the regional seismogenic area of Skyros, by studying the past seismicity of this area, and if any, to calculate the magnitude of the oncoming EQ. The period considered for the past seismic history of the area spans from 1988 to July 22$^{nd}$, 2001.
At a first approach, a magnitude of **M = 6.1R** was calculated. This value was based on the last "lock" period, detected, in the regional area of Skyros. The calculated magnitude equals Ms = 6.1R and fits the value, provided, by the Greek Seismological Observatory in Patras. This is shown in the following figure **(11)**.

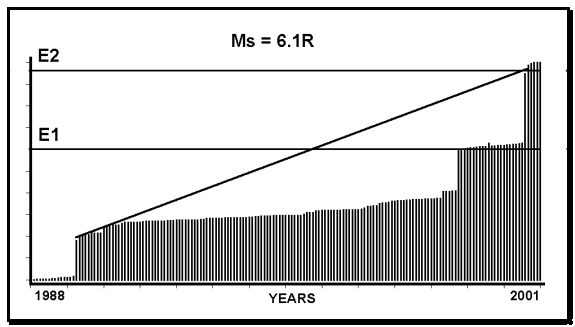

Fig. 11. Magnitude determination is shown, for the last lock period.

The normal energy out-flow rate at Skyros area, is indicated by the slope of the trend (solid line) that starts just after the first seismic event on the left side of the graph. In the next figure **(12)** a longer "lock" period is considered. The calculated magnitude equals **Ms = 6.3R** and fits the value provided by the Greek Seismic Observatory in Thessaloniki, Greece.

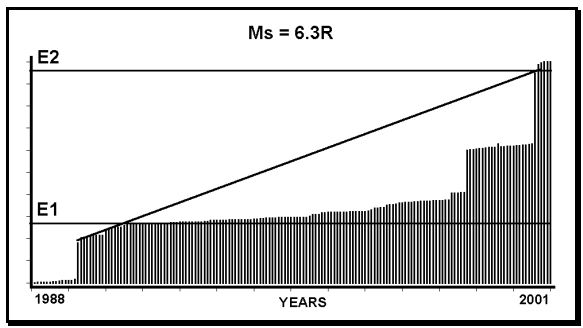

Fig. 12. Magnitude determination, when a longer lock period is considered.

Nevertheless, no matter what the actual, real magnitude of this EQ is, both solutions presented herein, are in very good agreement to the magnitude, calculated, by most seismological observatories.

**5.b. Northridge EQ, California, USA, 17/1/1994, M = 6.7 R**

The validity of the **LSEFM** methodology was tested, on the strong EQ of Northridge 17th January 1994 (M=6.7R, lat = 34.21, long = -118.54), too. In the following figure **(13)** the epicentral area of Northridge EQ is presented by a blue circle.

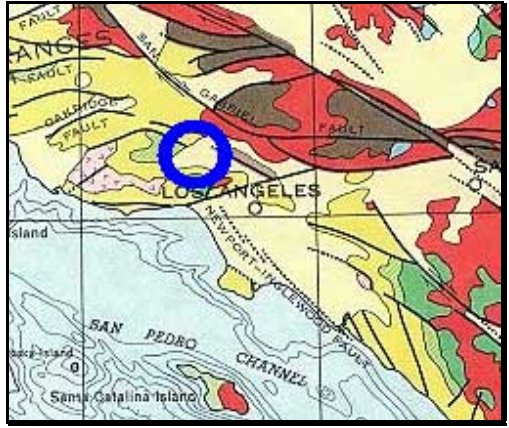

Fig. 13. Location map of Northridge EQ (blue circle) and the Los Angeles city (map after USGS).

Following the methodology of the **"LSEFM"**, the cumulative, seismic energy release graph was constructed for the period **1932** to **2004**. The sampling interval is **(1)** month. That is each **Y** value of this graph represents the total, seismic energy, released, over a month. The cumulative, seismic energy release was studied over a rectangular area around Los Angeles and is indicated by the coordinates:

Upper left: 34.75/-119.25  Upper right: 34.75/-117.25,

Lower left: 33.25/-119.25  Lower right: 33.25/-117.25.

The calculated, cumulative energy release graph is presented in figure **(14)**.

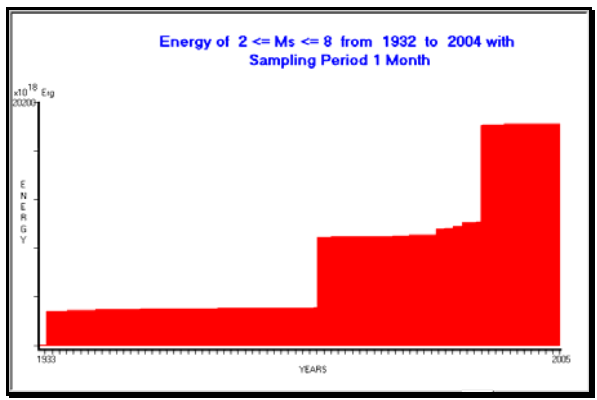

Fig. 14. Cumulative, seismic energy release, for the period 1932 - 2004, calculated, for the regional area of Northridge EQ.

The abrupt steps of the graph indicate the occurrence of strong EQs (in the past), in the studied frame.
From left to right the following three EQs are shown:

a. 1933, March 11th,    Lat. = 33.64  Long. = -117.97 Magnitude = 6.4R

b. 1970, February 9th, Lat. = 34.42  Long. = -118.37 Magnitude = 6.6R

c. 1994, January 17th, Lat. = 34.21  Long. = -118.54 Magnitude = 6.7R

What is evident, from the previous figure **(14)**, is the fact that the regional area of Northridge - Los Angeles remains in a "locked" state, for very long periods, acquiring seismic energy, through any suitable, physical, seismological mechanism, no matter what it is.
The accumulated, seismic energy is released, mainly, through strong EQs. Energy release through small-size seismicity does not modify, significantly, the appearance of the graph. The **LSEFM** method indicates indirectly, the "normal, cumulative energy release time-function", under unlocked conditions. Therefore, the next to come strong EQ magnitude can be calculated by taking into account the seismic history of the area under study.  This is presented in the following figure **(15)**.

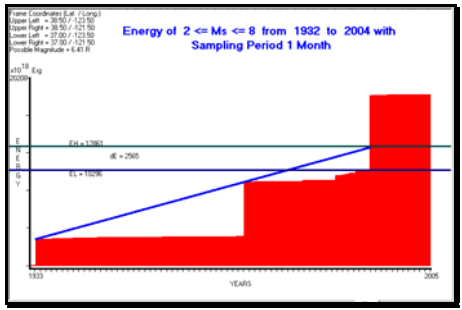

Fig. 15. Calculation of the magnitude of Northridge EQ is shown, by using the "**LSEFM**" methodology.

The seismological methods determined a magnitude for the EQ in Northridge of M = 6.7 R. The lithospheric, seismic energy flow model, which was applied to the regional area of Los Angeles, calculated its magnitude as M = 6.45 R. The normal, seismic energy flow rate is indicated by the two peaks of energy release, generated, from the two earlier, strong seismic events (1933, 1970). The difference in magnitude, between the two methods, is only dM = .29 R, not too bad at all. It must be pointed out, too, that the epicentral area and the timing of the EQ in Northridge, which were used in this test, were known in advance (a posteriori study).

### 5.c. Parkfield EQ, California, USA, 28$^{th}$ September 2004, Mw = 6.0 R.

The **LSEFM** method was applied to the regional area of Parkfield, in an attempt to calculate the magnitude of the earthquake on September 28th, 2004, Mw=6.0R. The details are as follows:

The adopted, seismogenic area is the one presented in the following figure **(16)**.

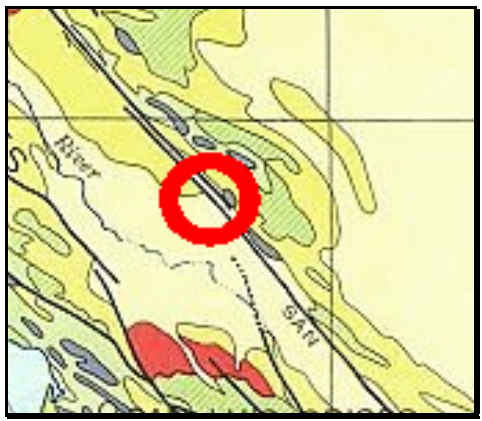

Fig. 16. Location – geological map of the earthquake in Parkfield (red circle) and used seismic, energy release frame location map (map after USGS).

The coordinates of the used frame are:

|  |  |  |  |
|---|---|---|---|
| Upper left: | 36.6 / -121.1 | Upper right: | 36.6 / -119.5 |
| Lower left: | 35.0 / -121.1 | Lower right: | 35.0 / -119.5 |

The calculated cumulative seismic energy release graph vs. time is shown in the following figure **(17)**.

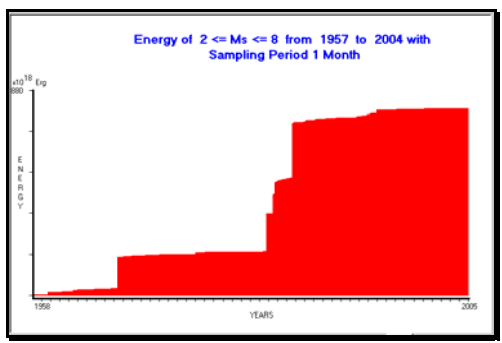

Fig. 17. Calculated cumulative seismic energy release is shown, for the period 1957 - 2004, for Parkfield area.

The calculated magnitude of Parkfield earthquake is Ms=5.94R **(**fig. **18),** compared to the Mw=6.0R, calculated, by seismological methods.

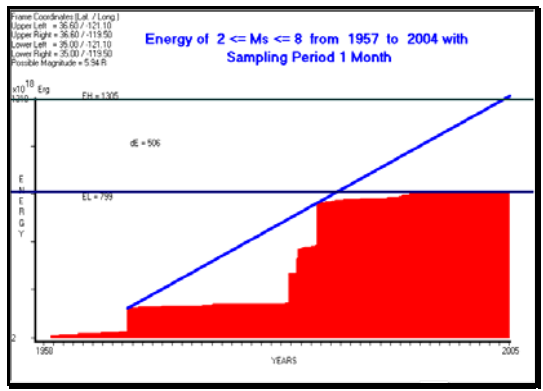

Fig. 18. Calculation of the magnitude of the strong EQ, in Parkfield area, that occurred on September 28[th], 2004.

## 6. Conclusions.

Summarizing all the above what can be said is as follows:
The seismogenic areas considered as open physical systems absorb energy for a long period of time and release it through their seismicity in the future. What is important is the fact that a) the study of their past seismicity suggests the way their energy is released under terms of a balanced state of energy mode, b) it can be calculated, indirectly, how much energy has been stored in it for a certain period which can be expressed as a single EQ when it is abruptly released.
The later indicates the way the magnitude calculation of a future large EQ can be made. This physical model was tested against real large earthquakes in Greece and USA with earthquake catalogs compiled by different seismological groups and the obtained results proved to be more than satisfactory in seismological terms.
In general, this methodology is a stand-alone procedure for the calculation of the maximum expected magnitude of a future earthquake in any seismogenic or not regional area even if there are no indications of any recent seismic activity in the under study area or any other seismic precursory data.
Prerequisites for the application of this methodology is the time of occurrence and the location of the future large EQ. The later can be determined by other methods (Thanassoulas, 2007).

## 7. References.